\documentclass[12pt,preprint]{emulateapj}

\newcommand{\kms}{km\,s${}^{-1}$}
\newcommand{\mymail}{oysteol@astro.uio.no}
\newcommand{\halpha}{\mbox{H\hspace{0.2ex}$\alpha$}} 
\begin{document}

\title{Measurements of plasma motions in dynamic fibrils}

\author{{\O}ystein Langangen}
\author{Luc Rouppe van der Voort\altaffilmark{1}}
\author{Yong Lin}

\affil{Institute of Theoretical Astrophysics, University of Oslo, 
P.O. Box 1029 Blindern, N-0315 Oslo, Norway}
\email{\mymail}

\altaffiltext{1}{also at Center of Mathematics for Applications, University of 
Oslo, P.O. Box 1053 Blindern, N-0316 Oslo, Norway}

\begin{abstract}
  We present a 40 minute time series of filtergrams from the red and
  the blue wing of the \halpha\ line in an active region near the
  solar disk center.
  From these filtergrams we construct both Dopplergrams and summed
  ``line center'' images. 
  Several dynamic fibrils (DFs) are identified in the summed
  images. The data is used to simultaneously measure the proper motion
  and the Doppler signals in DFs. For calibration of the Doppler
  signals we use spatially resolved spectrograms of a similar active
  region.
  Significant variations in the calibration constant for different
  solar features are observed, and only regions containing DFs have
  been used in order to reduce calibration errors.
  We find a coherent behavior of the Doppler velocity and the proper
  motion which clearly demonstrates that the evolution of DFs involve
  plasma motion.
  The Doppler velocities are found to be a factor 2--3 smaller than
  velocities derived form proper motions in the image plane. 
  The difference can be explained by the radiative processes involved,
  the Doppler velocity is a result of the local atmospheric velocity
  weighted with the response function. As a result the Doppler velocity
  originates from a wide range in heights in the atmosphere. 
  This is contrasted by the proper motion velocity which is measured
  from the sharply defined bright tops of the DFs and is therefore a very
  local velocity measure. The Doppler signal originates from well below the
  top of the DF.
  Finally we discuss how this difference together with the lacking
  spatial resolution of older observations have contributed to some of
  the confusion about the identity of DFs, spicules and mottles.
\end{abstract}

\section{Introduction}
\label{sec:intro}

The solar chromosphere owes its name to the reddish rim that appears
above the lunar limb during solar eclipses. 
This reddish color mostly stems from the Balmer {\halpha} spectral
line which makes this line one of the most important chromospheric
diagnostics.
Due to the highly dynamic state of the chromosphere and strong NLTE
effects, the line formation processes are still not yet fully
understood \citep[e.g.][]{radyn,2006Leenaarts}.
This is an important shortcoming in our interpretation tools which
makes \halpha\ observations traditionally difficult to interpret. 

Due to the highly fibrilar structure of the chromosphere \citep{1908Hale},
a strong influence from magnetic fields on the chromosphere has been 
suspected for about a century.

The most common of these fibrilar magnetic fine structures are the
jet-like structures known as spicules, mottles, and dynamic fibrils
(DFs). 
In short, spicules are traditionally observed at the limb, mottles on
disk in the quiet Sun, and DFs in active regions.

Whether or not these structures are manifestations of the same
phenomenon viewed at different angles have been the subject of a long
standing discussion
\citep[e.g.][]{1968Beckers,1992Grossman,1994Tsiropoula,1995Suematsu,
  2001Christo,2007Rouppe}.
One important argument against these structures being caused by the
same mechanism has been the difference in the measured absolute 
velocities \citep{1992Grossman}. 
Other authors have done direct measurements of mottles crossing 
the limb \citep{2001Christo}. They also state that since
both proper motions and Doppler motions are used in the comparisons,
systematic errors are probably introduced. Such errors might also be amplified
by the rather limited spatial resolution of some of the data sets used.

The detailed analysis of DFs has accelerated in recent years
\citep[e.g.][]{2004dePontieu,2006deWijn,2006Hansteen,2007dePontieu,
  2007Lars} due to major advances in both observational techniques and
simulation efforts.

One of the main conclusions from these studies is that DFs are driven
by magneto-acoustic shocks caused by p-mode oscillations and
convective flows leaking into the chromosphere.

In a recent study, \citet[][from now on paper 1]{2007bLangangen}
presented spectroscopic analysis of DFs as seen in one of the
Ca~{\small{II}}~IR lines.
Numerical analysis of the line formation process showed a much lower
DF velocity derived from Doppler measurements as
compared to the proper motion velocity. This was found to be due to
both the low formation height and the extensive width of the
contribution function of the Ca~{\small{II}}~IR line.
Furthermore, the DFs analyzed in paper 1 showed mass motion, 
thus ruling out any ionization/temperature wave as explanation model for 
DFs \citep[e.g.][and references therein]{2000Sterling}.  

With the advantage of well sampled spectral line profiles, the number
of analysed DFs in paper~1 was rather modest due to the limited
spatial coverage of the spectrograph slit.
With the current data we exploit the wide spatial coverage of a tunable
filter instrument, at the expense of limited spectral resolution. 
%

\begin{figure*}[!ht]
\includegraphics[width=\textwidth]{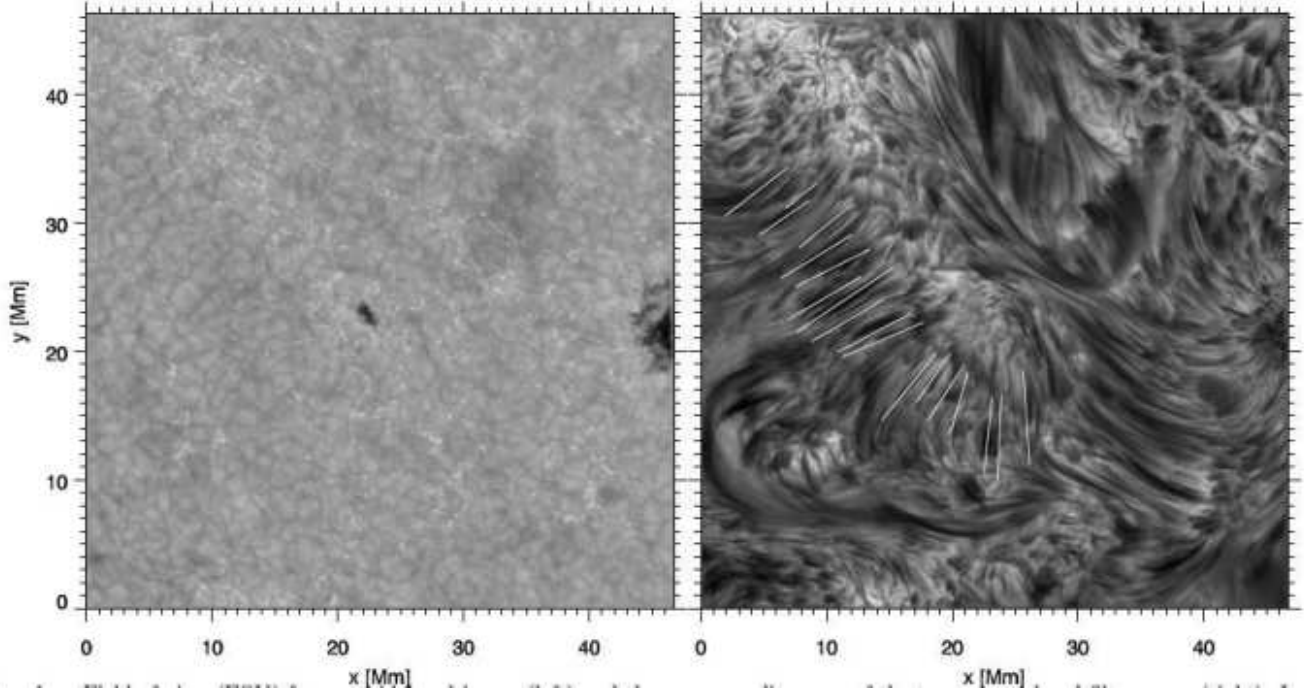}
\caption{
Field of view (FOV) for one wideband image (left) and the corresponding 
sum of the two narrow band filtergrams (right). 
In the narrow band image several fibrilar structures are seen, and some
DF axes are marked (solid white lines) for illustrative purposes.  }
\label{plotone}
\end{figure*}

In this paper we add to the understanding of these jet like structures
by analysis of Dopplergrams obtained in an active region close to
the disk center, hence the observed jet structures are commonly known as DFs.
In \S\ref{sec:obs} we describe the observational program and the 
instrumentation. The data reduction and the calibration method is explained 
in \S\ref{sec:datareduction}. In \S\ref{sec:Obsres} we present the results of
our measurements. We discuss our results
in \S\ref{Disc} and finally we summarize the results in \S\ref{Sum}. 

\section{Observing program and instrumentation}
\label{sec:obs}

The observations were obtained with the Swedish 1-m Solar Telescope
\citep[SST,][]{SST} on La Palma.  The degrading effects of
seeing were minimized by use of the SST adaptive optics system
\citep{SSTAO} and the Multi-Object Multi-Frame Blind
Deconvolution \citep[MOMFBD,][]{MOMFBD} image restoration
method. The Solar Optical Universal Polarimeter
\citep[SOUP,][]{1981Title} provided narrow band images in the
\halpha\ line (filter FWHM 12.8~pm). The optical setup is described in
detail in \citet{2007dePontieu}. Three fast Sarnoff CCD cameras,
operating at a frame rate of 37~frames~s$^{-1}$, were simultaneously
exposed by means of an optical chopper. One camera was operated as
SOUP camera, the other two cameras were positioned as a
phase-diversity pair on a beam that was split off from the main beam
before SOUP but behind the pre-filter (FWHM 0.8~nm). The latter
cameras provided wide-band photospheric reference images and operated
as MOMFBD anchor channel.

The target area (65\arcsec $\times$ 65\arcsec) was centered on a small
pore in NOAA AR10910, positioned at S11\degr, W11\degr (observing angle
$\theta=21\degr$, $\mu=\cos\,\theta=0.93$) on 23-Sep-2006, 
see Fig.\ref{plotone}. The time
series comprises 40~minutes, starting at 10:56:57 UT. The pixel scale
was 0.065\arcsec, 
the SST diffraction limit ($\lambda/D$) at 656.3~nm is 0.14\arcsec or
100~km.

A wavelength calibration was performed for SOUP to compensate for
several offsets like for example solar rotation. An \halpha\ profile
scan was obtained by stepping with 5~pm steps and averaging over a
relatively quiet region in the vicinity of the target area. This
profile scan was used to determine the line center shift in order to
provide correct positioning in the wings. The profile scan is shown in
Fig~\ref{plottwo}.

SOUP was alternating between the blue and red wing at $\pm30$~pm from
the \halpha\ line core. The total acquisition time for the two line
positions was 10.6~s: 1.1~s to record 40 exposures for each line
position and 8.4~s for changing line position. The line position
change time of SOUP is relatively long and is the limiting factor for
selecting the number of points to sample the spectral line profile. We
choose two line positions which is the absolute minimum for obtaining
Doppler information. Improving the spectral sampling would imply
unacceptably long acquisition times for which solar evolution changes
would dominate the resulting line profile.
Taking half a resolution element (approximately 60~km) as a limit, we
expect that motions in the image plane with velocities faster than
6~km~s$^{-1}$ cause false signals in the Dopplergrams.

\section{Data processing}
\label{sec:datareduction}

\subsection{Image post-processing}
\label{sec:imagepostprocessing}

All images from each SOUP cycle were jointly processed in a single
MOMFBD restoration, yielding three restored images: one wide band, one
\halpha\ red, and one \halpha\ blue wing image. Each restoration is
based on a total of 240 frames, 80 for each camera. The sequential
recording of the wing images and the long line position change time
imply that the seeing for the two positions is different. For the
construction of a Dopplergram from unprocessed images, one would
expect significant false signals due to misalignment. However, using
the wide band cameras as MOMFBD anchor channel, the restored \halpha\
wing images are guaranteed to be precisely aligned. The resulting
Dopplergram is therefore virtually free from misalignment errors.
This advantageous feature of restoring multiple objects with MOMFBD is
discussed in detail in \citet{MOMFBD}. Note however that
significant changes in the amount of blurring
can result in false signals after the subtraction process. 
For our data, we regard this as only a minor source of error
since the seeing was stable and homogeneous.

\begin{figure}[!ht]
\includegraphics[width=0.5\textwidth]{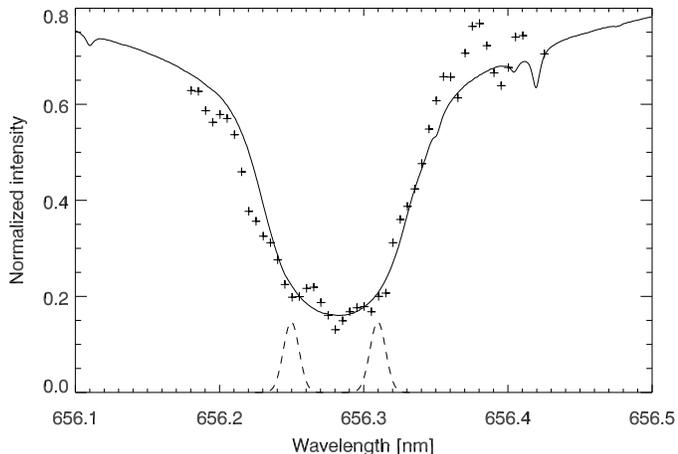}
\caption{ SOUP \halpha\ profile scan of a quiet region used for the
  line offset calibration (crosses).  A solar atlas profile is shown
  as reference (solid line).  Furthermore, the idealized filter
  profiles for the red and blue wing positions are plotted on an
  arbitrary scale (dashed lines).  }
\label{plottwo}
\end{figure}

Dopplergrams $D$ are constructed following the standard method
 \begin{equation}
D=\frac{B-R}{B+R}~.
\end{equation}
with $B$ and $R$ being the blue and red wing images respectively. 
The measured pixel values in the Dopplergrams will from now on be
referred to as the Doppler signal.
The sum of the wing images gives an image that is reminiscent of an
\halpha\ line core image. We use these summed images as substitutes
for line core images to measure the DF trajectories. 

The wide band and wing images form a time sequence with a cadence of
19.1~s. The frames in each time series are aligned and de-stretched
using the wide band images as reference to determine the (local)
offsets which were then applied to the wing images.
After alignment and de-stretching, Dopplergrams and summed images are
constructed from the wing images.

\subsection{Dopplergram calibration}
\label{sec:calibration}

\thispagestyle{empty}
\setlength{\voffset}{-14mm}
\begin{figure*}[!ht]
\includegraphics[width=\textwidth]{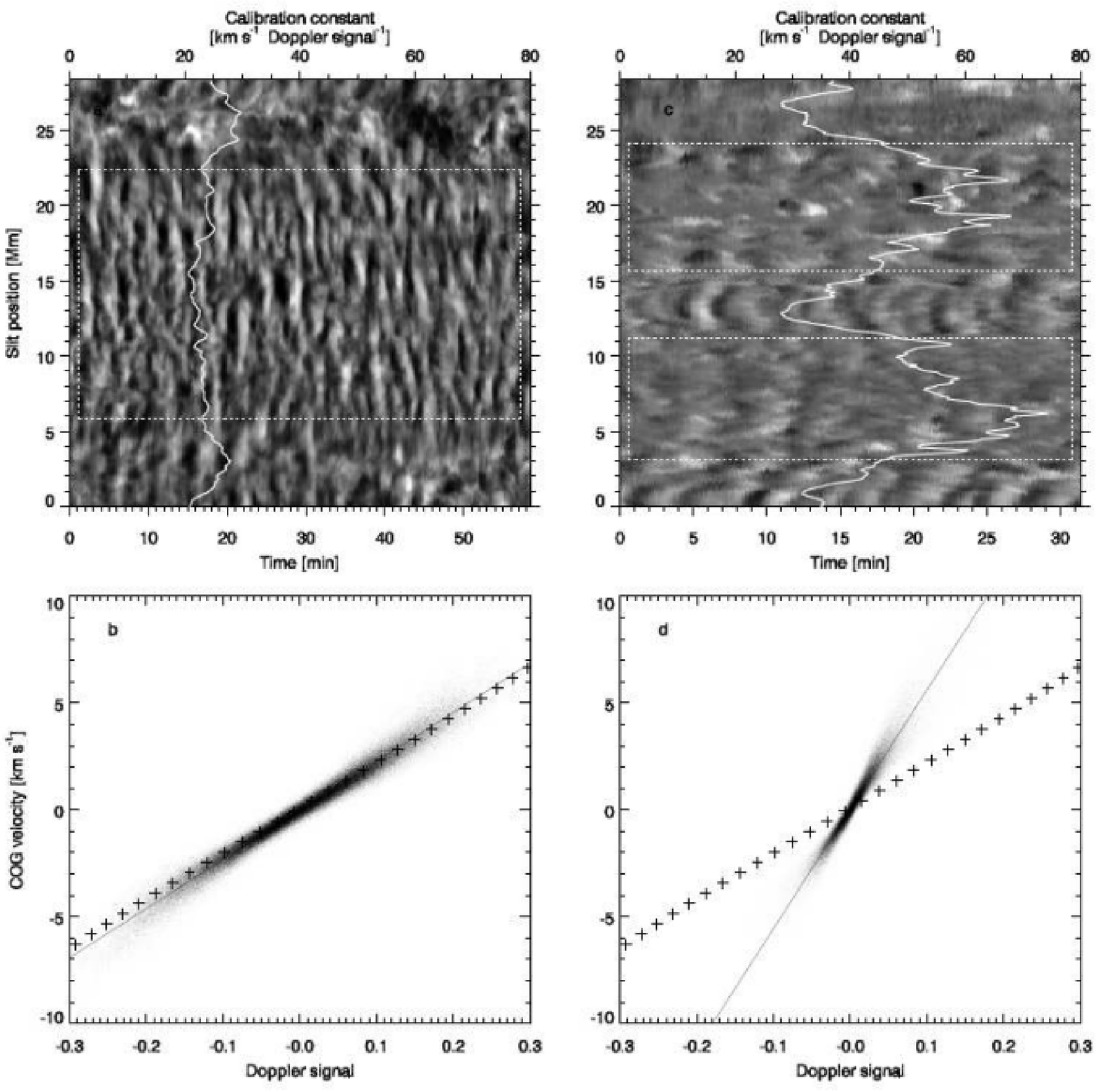}
\caption{
Relation between Doppler signal and velocity. The Doppler signals of a
quiet Sun spectrogram time series are shown in grey scale in panel a). White
corresponds to upflow and dark to downflow respectively. The white
solid line shows the spatially resolved
calibration constant derived from the temporal spread at that location
(top axis). The boxed area in panel a) covered
internetwork, the top and bottom parts mottles. 
The scatter plot of the Doppler signal and the COG velocity from the boxed 
area in panel a) are shown in panel b).
The least squares linear fit (solid line) as well as the artificially 
shifted atlas profile (crosses) are seen.
An active region timeseries is shown in panel c).
The two regions known to contain DFs are indicated by the two boxes.
The three excluded regions contain (lower and middle) running penumbral waves,
and (upper) long fibrils.
Again the calibration constant, assuming zero offset, is indicated (solid line).
The scatter-plot obtained from the two boxed regions in panel c is seen
in panel d).  The least squares linear fit (solid line) as well as the
artificially shifted atlas profile (crosses) are seen.  }
\label{plotthree}
\end{figure*}
\setlength{\voffset}{0mm}

\begin{figure}
\includegraphics[width=0.5\textwidth]{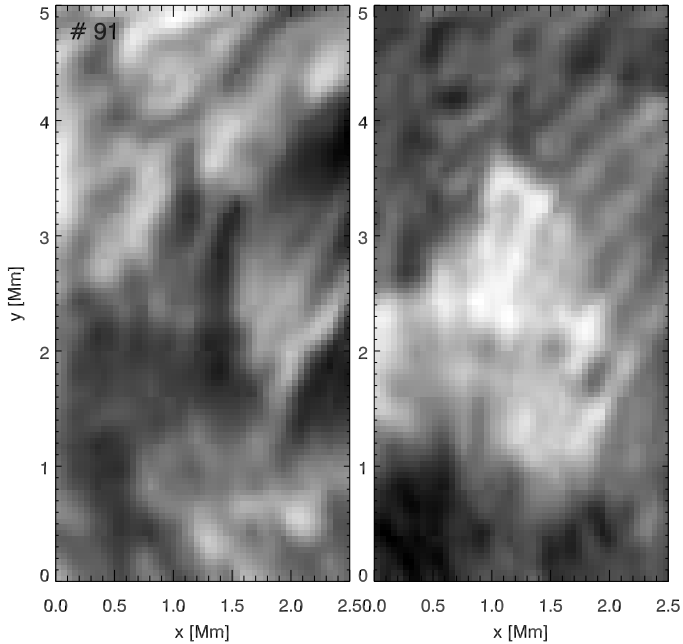}
\caption{ Example of an intensity image of a group of DFs (left
  panel), and the corresponding Doppler signal (right panel). In the
  Dopplergram, black and white signify upflow and downflow
  respectively. A movie of this DF is supplied in the electronic
  version of this paper.  }
\label{plotfour}
\end{figure}

\begin{figure*}
\includegraphics[width=\textwidth]{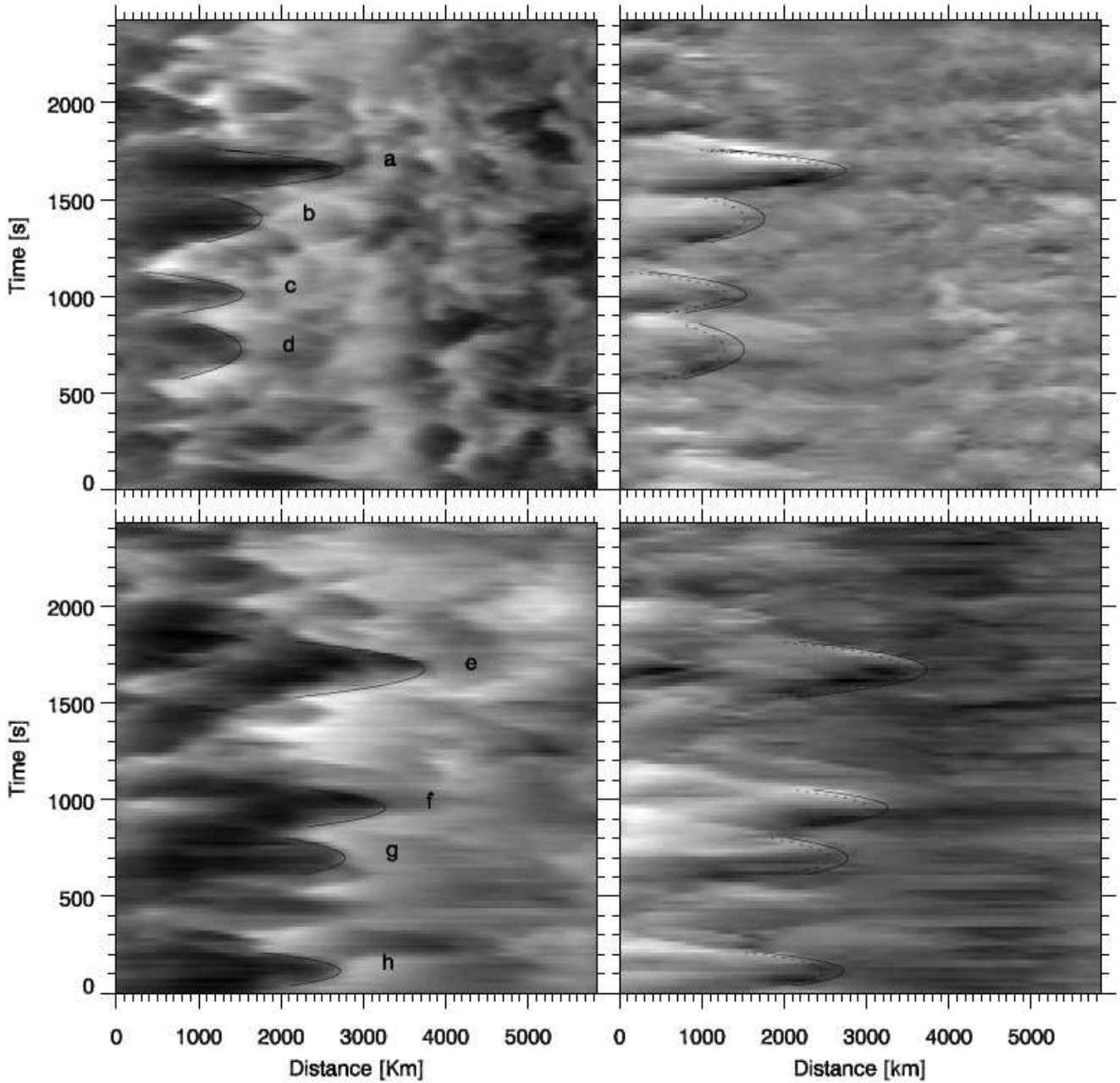}
\caption{ Two illustrative examples of summed x--t plots are seen in
  the upper and lower left panels. The corresponding Dopplergrams are
  seen in the two panels to the right. The parabolas fitted in the
  intensity image (solid) and the parabolas used in the Doppler images
  (dashed) are seen. Each parabola is marked with a letter which
  corresponds to the individual DF plots presented in
  Fig.\ref{plotsix}.  }
\label{plotfive}
\end{figure*}
\begin{figure}
\includegraphics[width=0.5\textwidth]{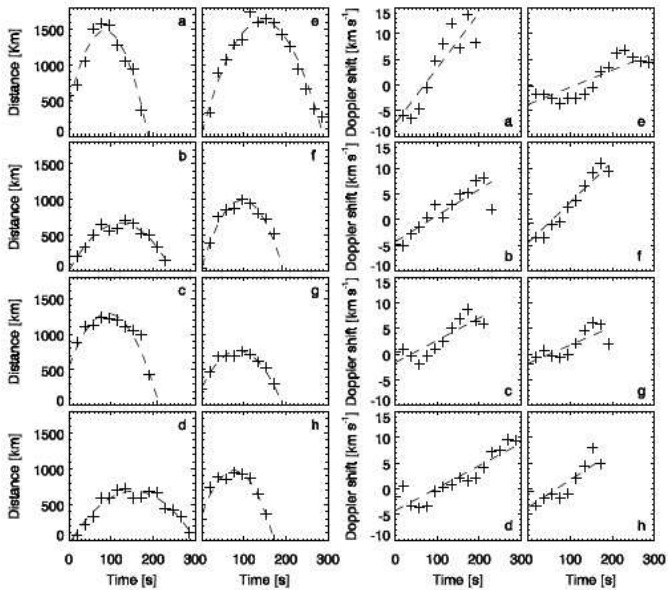}
\caption{ 
 For illustration of our observations we have plotted eight examples
of measured DFs. The eight panels to the left show the intensity fitted 
parabolas, while the eight panels to the right show the corresponding 
Doppler shifts. The letters in each panel makes it possible to identify 
the corresponding DFs seen in Fig.\ref{plotfive}.
}
\label{plotsix}
\end{figure}
\begin{figure}
\includegraphics[width=0.5\textwidth]{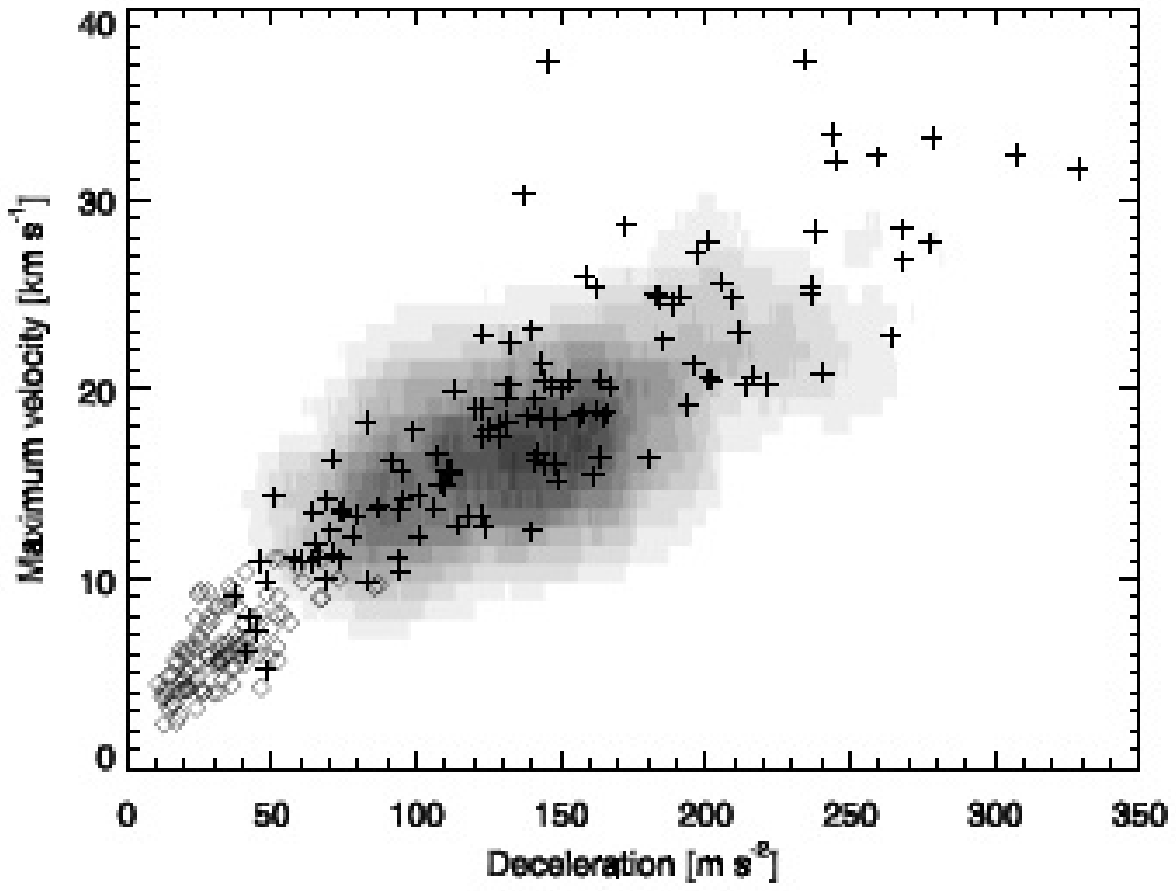}
\caption{Scatter plot of the decelerations and maximum velocities.
The earlier reported values \citet{2006Hansteen} is shown (scatter cloud).
The image data results (crosses) and the Doppler data results(diamonds)
are over plotted.}
\label{plotseven}
\end{figure}

The Dopplergrams were calibrated in order to interpret the signals 
as Doppler velocities. 
The main concern with such a calibration is the large variation in the
shape of the {\halpha} line profiles from different regions of the Sun.

As a first order approach we use the {\halpha} atlas profile from the
Kitt Peak solar spectral atlas \citep{1987Brault}.
Artificial Doppler shifts are applied to the atlas profile, and the
Doppler signals are measured, using the same procedures as for the 
observations.
For low velocities, there is a linear correlation between the velocity
and Doppler signal, see the bottom panels of Fig.~\ref{plotthree}. 
For velocities approaching 13.7~{\kms}, the velocity corresponding to a
Doppler shift of 30~pm, the correlation starts to deviate from linear.
Only for velocities that follow the linear correlation with the
Doppler signal, a simple calibration constant can be used.
We estimate that for the line positions used, $\pm$30~pm, this
corresponds to a velocity of about 9~{\kms}, or a Doppler signal of
$\pm$~0.4.

The atlas profile is derived from averaging observations from disk
center quiet Sun. The calibration constant derived from this profile
might not necessarily apply to Doppler observations from other solar
structures. 
To investigate the effect of variations in the shape of {\halpha} line
profiles on Doppler signals we analyze spectrogram timeseries of both
quiet Sun and an active region.

The quiet sun spectral timeseries was obtained on the same day,
04-May-2006, as the active region time series described in paper 1.
This timeseries covers about an hour of a fairly quiet internetwork
region at disk center, obtained during excellent seeing conditions.
As for the spectral atlas, we measure both the Doppler velocity and
the Doppler signal, but now we measure this in the resolved spectra.
The observed {\halpha} line profiles are generally very wide and flat
and have therefore rather low Doppler sensitivity.
To cope with these problems the Doppler velocity was measured using
the center--of--gravity (COG) shift of the $4\%$ lowest intensity of
the line. 
This method gives more weight to the Doppler velocities from
the near wings, similar to the Doppler signal measurement. 
The resulting correlation between the Doppler signal and the Doppler
shifts is shown in panel b in Fig.~\ref{plotthree}.  
It is clear that the result from the quiet Sun series is very similar
to the relation found for shifting the atlas profile. 
This is because the width of the resolved line profiles in this region
do not differ substantially from the atlas profile.

The active region spectral timeseries is described in detail in
paper~1.  This series covers a plage region containing several pores.
The viewing angle ($\mu =0.96$) is approximately the same as the angle
in the Dopplergrams from the present paper, hence any systematic
errors introduced by differences in the viewing angle are small.
As above, both the Doppler velocity and the Doppler signal are
measured in this timeseries.
Significant differences compared with the spectral atlas calibration
are found, see Fig.\ref{plotthree} panels c--d.
The calibration constant is in general higher in the active region,
typically by a factor two. 
It is, however, clear from the spatially resolved calibration
constants (panel c in Fig.\ref{plotthree}), that the variation is
quite large.
This variation is caused by the different spatial structures covered
by the slit.
The lower and middle part of panel c) covered pores hosting running
penumbral waves, the upper part covered low lying fibrils. In these
regions, smaller calibration constants are found.
In the boxed areas of panel c), there is also some variation, but the
lowest values are always substantially higher than for the quiet sun.
These areas hosted several dynamic fibrils. 
Similar results were found in other time series, but with slightly
lower calibration constants. We attribute this to the seeing
conditions, which were less favourable.

We conclude that observed Doppler signals are very dependent on the
shape of the {\halpha} line profile.
The shape of the line profile varies significantly for different solar
structures and the calibration of Dopplergrams has to be done with care.
The use of spatially resolved spectra is highly desirable, but even
then the variations can be quite high.
For the remainder of this paper we adapt a calibration constant of
55~{\kms}\,Doppler signal${}^{-1}$ 
, the mean value from the two boxes in panel c in Fig.~\ref{plotthree}.
We argue that the calibration constant from these boxes is the optimal
choice for our Dopplergrams since we are interested in DFs. Numerous
DFs were found in the region covered by the boxes in panel c). 
The standard deviation of the calibration constants in these resolved
spectra is 7~{\kms}. Note that this error should be multiplied
by the Doppler signal, which gives a typical error of about 1--1.5
km\,s${}^{-1}$.

\subsection{Trajectory measurement method}
\label{sec:method}

For the measurement of the line of sight (LOS) Doppler velocity and
the proper motion in the image plane, we follow the methods of
\citet{2006Hansteen} and \citet{2007dePontieu}.
The DFs are identified by visual inspection of the summed intensity
time series, and DF axes are manually defined. We see that individual
DFs follow approximately straight trajectories.
The right panel in Fig.~\ref{plotone} shows examples of such axes, see also
Fig.\ref{plotfour} for a more detailed view of a DF.
Data along these axes is extracted from both the summed and Doppler
data 'cubes' for further analysis.
From the extracted data we construct x--t plots for measuring DF
trajectories.  
Some examples of x--t plots showing both the intensity and the Doppler
signal are seen in Fig.\ref{plotfive}.

The trajectory is defined by the position of the maximum change in the 
intensity between the top of the DF and the background, and a 
parabola is fitted to these data points.
These parabolic fits give a good description of the temporal
evolution of the length of the DFs, see Fig.\ref{plotsix}.
The velocity in the image plane is given by the time derivative of the
fitted parabola.

The Dopplersignal is extracted from the Dopplergrams, using the
positions given by the parabolic fits obtained from the filtergrams.
For a more robust measurement of the Doppler signal, we extract the
Dopplersignal 5 pixels ($235$ km) below these points along the DF
axis, for further discussion, see \S~\ref{sec:errors}.
The Doppler data are fitted with a linear least square fit, which
gives a good description of the time evolution of the velocity, see
Fig.\ref{plotsix}.
The maximum velocity and deceleration is derived from the linear fit.

\subsection{Error estimation}
\label{sec:errors}

In this section we summarize the different sources of error and
discuss to what extent these affect our measurements of DF
trajectories.
The uncertainty in the line center calibration through the SOUP
profile scan affects the wavelength positioning of the SOUP filter
(see Fig.~\ref{plottwo}).
This could introduce an offset in the Doppler velocity which affects
the determination of the maximum Doppler velocity. 
To get a reasonable estimate on the error in this calibration, we
have shifted the atlas profile relative to the observational points.
Visual inspection of these shifted fits gives us an  
estimate of the error of the order of $\pm1$~km\,s${}^{-1}$.

The long tuning time, 8.4~s, between the two spectral positions
introduces errors due to changes in the seeing and solar evolution. In
\S~\ref{sec:imagepostprocessing}, we argue that the error from
seeing is minimal due to the favorable conditions. Furthermore, the
MOMFBD method guarantees precise alignment of the SOUP images. In
\S~\ref{sec:obs}, we estimate that motions in the image plane that
are faster than $\approx$6~{\kms} cause false signals. This is a
serious concern since DFs have maximum velocities that are far greater
than 6~{\kms}. For the summed images, this means that one would expect
the top end for the DFs to be less well defined during the beginning
and end of their lifetime, when the velocity is largest. However, for
the measurement of the DF trajectory this is of lesser concern since
most of the measured points are found during the period close to
maximum height when the velocity is lowest. The deceleration and
maximum velocity are determined from the parabolic fit.
Also in the Dopplergrams, the changes due to solar evolution mostly
effect the sharp edge of the DFs and significant false signals can be
expected during the periods close to maximum velocity. This is why we
choose to measure the Doppler values a few pixels lower than the top
of the DF. Since the DFs have considerable linear extent that appear
to be moving rather coherently, we estimate that this way we reduce the
effect of solar evolution on the measured maximum Doppler velocity.
Like for the summed images, the deceleration is determined from a fit
where most of the fitted points are found when the velocity is
lowest. 
In the same manner, using the linear fit reduces the impact of the
increasing error for larger velocities, when the velocity approaches
values corresponding to the $\pm30$~pm filter offset.

As mentioned in \S\ref{sec:calibration} the calibration constant 
in the region where DFs
are found varies, this can probably be attributed to variations in the
magnetic field topology. This introduces an error
of about 1--1.5 km\,s${}^{-1}$. Combining all the errors in the Doppler
velocity measurements gives an error of approximately 2~km\,s${}^{-1}$.
The proper motion velocities has an error of about 1~km\,s${}^{-1}$.

\section{Results}
\label{sec:Obsres}

A total of 124 DFs are identified throughout the 40~minute time
series. 
One example of a group of DFs is seen in Fig.\ref{plotfour}.
These DFs are seen to move in a semi-coherent manner, as illustrated
by the Dopplergram which displays downflow signal for the whole patch
of DFs.
Such semi-coherent behavior of groups of DFs is also described in
\citet{2007dePontieu}.

\begin{table}[!ht]
\caption{Statistical properties of DFs.\label{table1}}
\begin{tabular}{rrrrr}
\tableline
\tableline
Life time& Proper vel.& Proper dec.& LOS vel.&LOS dec. \\
{[s]}&[km\,s${}^{-1}$]&[m\,s${}^{-2}$]&[km\,s${}^{-1}$]&[m\,s${}^{-2}$] \\
\tableline
$258\pm[56]$ &$18.6\pm[6.6]$ &$142\pm[64]$ &$6.4\pm[2.1]$ &$33.7\pm[15.6]$ \\ 
\tableline
\end{tabular}
\end{table}

Using the methods described in \S~\ref{sec:method} we extract
maximum velocities, decelerations, and lifetimes of the 124 DFs, both
in the image plane and along the LOS.
For the Doppler measurements we include only those with an increasing
velocity with time, and with a change in velocity of more than $3$
km\,s${}^{-1}$ over the DFs lifetime.
Using these criteria we remove 18 DFs, which means that more than
$85$\% of the DFs display clear Doppler shifts together with proper
motion.
The statistical properties extracted for these DFs (124 based
on proper motion and 106 based on Doppler motion) are presented in 
Table.\ref{table1}. 
The mean values are showed together with the corresponding
standard deviations in brackets.

The deceleration and the maximum velocity show a strong correlation both
in the image data and the Doppler data, see Fig.\ref{plotseven}.

\section{Discussion}\label{Disc}

The correlation between the maximum velocities and decelerations found
from the proper motion measurements is similar to the correlation
found by \citet{2006Hansteen} and \citet{2007dePontieu}.
This is illustrated in Fig.~\ref{plotseven} by the grey-scaled cloud
shown in the background.
This correlation between the deceleration and the maximum velocity is
known to be the signature of shock waves being the driving mechanism of
DFs \citep{2006Hansteen,2007dePontieu,2007Lars}.

Further support for this driving mechanism comes from the fact that we
find a coherent behavior between the evolution of the Doppler signal
and the proper motion for a large fraction of the DFs. 
This is a strong indication that there is actual plasma motion
occurring during the life time of DFs. 
This supports the findings of paper~1, but based on a much larger
sample.

The Doppler measurements show a similar correlation as for the proper
motion, but with much lower absolute values for both the decelerations
and maximum velocities.
One possible explanation for these lower values could be high
inclination angles of the DF trajectories with the LOS.
One could expect to be able to derive the full trajectory vector
by combining the two measured deceleration components. 
This naive method would give very high inclination angles, typically
$75^\circ$. 
We know, however, that this can not be the true inclination angle,
since the Doppler velocity is a result of the local atmospheric
velocity weighted with the response function to velocity over an
extended height. 
In contrast, the measured proper motion is very local due to the high
contrast boundary between the top of the fibril and the surroundings.
Combining these two measurements leads to highly overestimated
inclination angles.
The difference in absolute values must be considered in the context of
the results from Paper~1.
The lower Doppler velocities found from the Ca~{\small{II}}~IR line
was explained by a combination of lower formation height and extended
formation range. 
This is probably also the case for the \halpha\ line, but
the formation height usually extends over a larger height range as compared
to the Ca~{\small{II}}~IR line.

The lacking Doppler shifts in ${\sim}15$\% of the DFs can either
be caused by very high inclination angles, 
or their driving mechanism is fundamentally different and the
evolution of these DFs is not a result of mass motion.  
We believe that high inclination angles combined with the
uncertainties in the measurements is a more plausible explanation for
the lacking Doppler signals.
The identification method of DFs introduces a bias toward the more
inclined DFs.
There are a number of suggestive cases where DFs are visible in the
Dopplergrams but no clear signature can be seen in the corresponding
intensity images.
We refrain from measuring these DFs since this will complicate a
comparison with other data sets.
Furthermore, the identification of these DFs is not objective and we
expect the measurement errors to be unacceptably high since low
inclination angles would lead to potentially high Doppler velocities.
Due to the lacking spectral sampling this could lead to strong 
saturation effects in the measured Doppler velocities.

\subsection{Spicules, mottles and fibrils}

The identification of the disk counterpart of spicules was already an
important question forty years ago \citep[e.g][]{1968Beckers}. 
One of the main problems was to reconcile the velocities measured in
spicules with those measured in mottles. 
This problem was also the main concern of \citet{1992Grossman}.
They conclude that since the velocities are much larger in spicules
than in mottles, the two could not be the same structure seen at
different viewing angles.
They, however, admit that the seeing might impair their results if the
structures observed were smaller than $1\arcsec{}$. 
Later studies of mottles and spicule properties lead to the conclusion
that spicules and mottles are in fact the same feature seen at
different angles \citep{1993Tsiropoula,1994Tsiropoula}.
\citet{1994Tsiropoula} showed, using cloud modelling, that the proper
motion and the cloud velocities were consistent.
In a more recent work, \citet{2001Christo} use a limb darkening
correction method to directly observe mottles crossing the
limb. 
They argue that the main reason for earlier confusion is caused by the
lacking spatial resolution of the observations.

In our study it is clear that the Doppler signal originates
from spatially resolved structures.
The excellent quality of the observations largely removes the errors
due to lacking spatial resolution.
Assuming reasonable inclination angles, i.e. the mean angles being not
very large nor very small, we can conclude that the Doppler velocities
are typically a factor of $\sim 2$--$3$ smaller than the corresponding
proper motion.
A similar, but larger difference was reported by
\citet{1994Tsiropoula}, we believe that this difference can be
attributed to worse spatial resolution.

As discussed above, radiative transfer processes are the fundamental
reason for the Doppler velocities being lower than the proper motion.
We argue that the fundamental differences between Doppler and proper
motion velocities that we find for DFs, also are valid for similar 
measurements on spicules and mottles. 
Besides the arguments of lacking spatial resolution, we believe that
this difference was an important contributor to the earlier confusion
about the unification of mottles and spicules.
Similar work on spatially resolved limb spicules is needed to finally
settle this discussion.

\section{Summary}\label{Sum}

With the SST we have obtained a 40 minute timeseries of an
active region observed in two line positions in the {\halpha}
spectral line.  For the first time, proper motion and Doppler
velocity of DFs can be simultaneously measured in spatially 
resolved observations.

We find that most DFs, about 85\% of our sample, show both co-temporal
and co-spatial Doppler motion and proper motion. The coherent behavior
of the two velocity components shows that the evolution of DFs involve
real plasma motion.

Both the proper motion and the Doppler measurements show a
strong correlation between the maximum velocity and the
deceleration.  This is in agreement with earlier findings and
supports the theory that DFs are driven by magneto acoustic shocks.

We derive significantly lower values for the deceleration and maximum
velocity from the Doppler measurements. We argue that this can be
explained by the height extension of the response function.  The
Doppler velocity is a result of the atmospheric velocity weighted with
the response function over an extended height. The proper motion
velocity is derived from the large contrast between the top of the DF
and the background which is a very local measure.

Using high resolution spectrograms we have demonstrated the
importance of a rigid calibration method for Dopplergrams.

\acknowledgements
Mats Carlsson is thanked for improving the manuscript.
The Norwegian Research Council supports Luc Rouppe van der Voort 
through grant 146467/420 and Yong Lin through grant FRINAT 171012.
The Swedish 1-m Solar Telescope is operated on the island of La Palma
by the Institute for Solar Physics of the Royal Swedish Academy of
Sciences in the Spanish Observatorio del Roque de los Muchachos of the
Instituto de Astrof{\'\i}sica de Canarias.
This research has made use of NASA's Astrophysics Data System.
\bibliographystyle{apj}
\bibliography{paper3}

\end{document}